\documentclass[aps,twocolumn,showpacs,preprintnumbers,amssymb]{revtex4}

\newcommand{\be}{\begin{eqnarray}}
\newcommand{\ee}{\end{eqnarray}}

\newcommand{\eins}{\mbox{$1 \hspace{-1.0mm}  {\bf l}$}}
\def\bea{\begin{eqnarray}}
\def\eea{\end{eqnarray}}
\def\C{\hbox{$\mit I$\kern-.7em$\mit C$}}
\def\N{\hbox{$\mit I$\kern-.3em$\mit N$}}

\def\l{\langle}
\def\r{\rangle}

\def\tr{{\rm tr}}

\usepackage{dcolumn} 
\usepackage{bm} 

\usepackage{epsfig}

\begin{document}

\title{Quantum random walks in optical lattices}

\author{W. D\"{u}r$^1$, R. Raussendorf$^1$, V. M. Kendon$^2$ and H.-J. Briegel$^1$}

\affiliation{
$^1$ Sektion Physik, Ludwig-Maximilians-Universit\"at M\"unchen, Theresienstr.\ 37, D-80333 M\"unchen, Germany.\\
$^2$ Optics Section, Blackett Laboratory, Imperial College, London, SW7 2BW, United Kingdom.}


\date{\today}

\begin{abstract}
We propose an experimental realization of discrete quantum random walks using 
neutral atoms trapped in optical lattices. The random walk is 
taking place in position space and 
experimental implementation with present day technology ---even using existing 
set--ups--- seems feasible. We analyze the influence of possible imperfections 
in the experiment and investigate the transition from a quantum random walk to 
the classical random walk for increasing errors and decoherence.  
\end{abstract}


\maketitle

\section{Introduction}

The increasing effort to investigate theoretically and experimentally the 
possibility to construct and build an universal quantum computer is mainly 
motivated by the expectation that quantum computers offer a (possibly exponential) 
speed--up over classical computers. Despite the two celebrated milestones of 
Shor's factoring algorithm \cite{Sh96} and Grover's database search algorithm 
\cite{Gr98} ---which both offer a speedup over their best (known) classical 
counterpart--- no constructive way to generate efficient quantum algorithms is currently known.
One possible direction of research is the adaption of known 
classical algorithms to the quantum mechanical case.

Random walks on graphs play an essential role in various fields of natural 
science, ranging from astronomy, solid--state physics, polymer chemistry  and 
biology to mathematics and computer science \cite{Ba70}. In particular, markov 
chain simulation has emerged as a powerful algorithmic tool and many classical 
algorithms are based on random walks. It is possible and hoped that quantum 
random walks allow in a similar way for a constructive search for new quantum 
algorithms. This justifies the increasing effort in the investigation of quantum 
random walks by several groups 
\cite{Ah93,Fa98,Ah00,Am01,Ma02,walks,Ke02,Ya02,Tr02,Ba02,Sa02}. Different behavior of 
the quantum random walk ---as compared to the classical one--- have been 
reported under various circumstances. For instance, a very promising feature of 
a quantum random walk on a hypercube, namely an exponentially faster hitting 
time as compared to a classical random walk, has been very recently 
found (numerically) by Yamasaki et al. \cite{Ya02} and (analytically) by Kempe \cite{Ke02}.

In this paper, we consider the simplest and best--studied version of a random 
walk, namely the discrete Hadamard walk on a line or a circle, first studied by 
\cite{Ah93}. We propose an experimental implementation of the quantum random 
walk using neutral atoms trapped in an optical lattice. In contrast to the 
recently proposed implementations using ion traps put forward by Travaglione and 
Milburn \cite{Tr02} and microwave cavities put forward by Sanders et al. 
\cite{Sa02}, in our proposal the random walk is taking place in position space 
and several hundred steps may be implementable even with present day technology.  

The paper is organized as follows. In Sec. \ref{classical}, we compare 
features of the classical and quantum random walk on the line and introduce some 
basic notation. Sec. \ref{implementG} provides a description of the physical 
set--up using optical lattices and implementations of the Hadamard walk on a line 
using this set--up. Possible imperfections and their influence 
on the quantum features of the walk are discussed in Sec. \ref{imperfect}. We 
summarize and conclude in Sec. \ref{summary}.

\section{Classical vs. quantum random walks}\label{classical}

\subsection{Classical random walk on a line}\label{classicalwalk}

Consider an infinite line with allowed (integer) positions $x_k \equiv k$, $k \in {\mathbb Z}$ and a 
particle which is initially located at position $x_0=0$. We consider a 
step--wise evolution in such a way that at each step, the particle moves with 
probability $1/2$ one step to the left, $x(n)=x(n-1)-1$, and with 
probability $1/2$ to the right, $x(n)=x(n-1)+1$. After $n$ steps, the 
probability $p_{\rm classical}(n,k)$ to find the particle at position $x_k$ is given by
\be
p_{\rm classical}(n,k)=\frac{1}{2^n}\left( \begin{array}{c} n  \\ \frac{k+n}{2}  \end{array} \right)
\ee
Note that if $n$ is even [odd], only even [odd] positions are occupied. The 
standard deviation of the distribution is $\sqrt{n}$, which implies a spreading time 
proportional to $\sqrt{n}$. The probability to observe a particle at 
distance of order $n$ from the origin decreases exponentially with $n$ and is zero $\forall n_0$, $n_0 > n$.

\subsection{Quantum random walk on a line}

A quantum mechanical analogy for the classical random walk would be a particle 
whose state evolves at each step into a coherent superposition of moving one step 
to the right and one step to the left. One readily finds \cite{Me96} that unitarity of the 
evolution implies that one has to consider a particle with internal degrees of 
freedom to achieve this aim. We consider a particle with two internal degrees of 
freedom, which can move on an infinite line with integer positions \cite{Ah00,Am01}. The 
corresponding Hilbert space ${\cal H}={\cal H}_I\otimes {\cal H}_X$ is given by 
${\cal H}_I=\C^2$, the internal state of the particle, and ${\cal H}_X=\C^\infty$, 
 the position space with basis states $|k\r_X \in {\cal H}_X$,  $-\infty \leq k \leq \infty$, corresponding to the particle located at the $k^{\rm th}$ lattice site..

The internal state of the particle, $|0\r, |1\r$ determines the direction of the 
particle movement. If the internal state is $|0\r$, the particle moves 
to the left, while it moves to the right if the internal state is $|1\r$. This 
operation is described by the unitary controlled--shift operation,
\be
\label{cshift}
S=|0\r\l0|\otimes \sum_k |k-1\r\l k| + |1\r\l1|\otimes \sum_k |k+1\r\l k|, 
\ee
i.e. $S|0\r_I \otimes |k\r_X = |0\r_I\otimes |k-1\r_X$ and $S|1\r_I \otimes |k\r_X = |1\r_I\otimes |k+1\r_X$.

We also introduce the Hadamard operation,
\be	
\label{Had}
H=\frac{1}{\sqrt{2}} \left( \begin{array}{cc} 1 & 1 \\1 &-1 \end{array} \right)
\ee
which acts on Hilbert space ${\cal H}_I$ such that $H|0\r=1/\sqrt{2}(|0\r + 
|1\r)$,  $H|1\r=1/\sqrt{2}(|0\r - |1\r)$. The particle is initially prepared in 
state $|\psi_0\r=1/\sqrt{2}(|0\r_I+i|1\r_I)\otimes |0\r_X$. Each step of the quantum random walk 
---which is also called Hadamard walk--- consists of applying the Hadamard operation, $H\otimes\eins$, followed by the 
controlled--shift operation $S$. Let $|\psi_n\r=(SH)^n|\psi_0\r$ be the state of the system 
after $n$ steps. The probability $p(n,k)$ to observe a particle at position $k$ 
after $n$ steps is given by 
\be
\label{probHad}
p(n,k)=\tr_X(|k\r_X\l k| \tr_I(|\psi_n\r\l \psi_n|)),
\ee  
and may be compared to the probability distribution $p_{\rm classical}(n)$ of the classical random walk. The 
probability distribution $p(n)$ has been analyzed in detail in Ref. \cite{Am01}.
 While $p_{\rm classical}(n)$ is given by a binomial distribution ---which is for large $n$ well approximated by a 
gaussian---, no such simple form exists for $p(n)$. The standard deviation of the distribution 
$p_{\rm classical}(n)$ is $\sqrt{n}$, while $p(n)$ is almost uniformly distributed in the 
interval $(-n/\sqrt{2},n/\sqrt{2})$ and the standard deviation is linear in $n$. 
This implies that the spreading time for a particle goes like $\sqrt{n}$ for the 
classical random walk, while in the quantum random walk it scales {\it linearly} 
with $n$. This provides an essential different behavior of the quantum random walk 
that follows from the possibility of interference in the 
quantum mechanical case. 

In a similar way, the (quantum) random walk on a circle is defined using a position space ${\cal 
H}_X=\C^N$ with periodic boundary conditions, i.e. $|k\r_X=|k {\rm mod}(N)\r_X$ 
for some finite $N$. Also in this case, a quadratic speed--up of the quantum 
random walk compared to the classical random walk is found in the spreading time 
of the particle \cite{Ah00}. Walks on general graphs can be defined in a similar 
way \cite{Ah00}.

\section{Implementation in optical lattices}\label{implementG}

In this section, we discuss possible implementations of the quantum random 
(Hadamard) walk on a line or on a circle using neutral atoms trapped in periodic 
optical potentials (for a review see e.g. \cite{Je01,Gr99}).

\subsection{Physical set--up}

We consider two identical one--dimensional optical lattices, each of them 
trapping one of the internal states $|0\rangle,|1\r$ of a neutral atom. For 
example, one may use alkali atoms with a nuclear spin equal to 3/2 ($^{87}$Rb, 
$^{23}$Na), and choose the hyperfine structure states $|F=1,m_f=1\rangle$ 
[$|F=2,m_f=2\rangle$] to represent $|0\rangle [|1\rangle]$ respectively. 

Each lattice consists of a periodic optical potential with period $d$. The 
optical potentials are formed by the standing waves resulting from two 
counter--propagating traveling waves with the electric fields forming an angle 
of $2\theta$, the so called lin$\angle$lin configuration. By changing $\theta$, 
the right and left circular polarized components $\sigma^\pm$ of the standing waves forming the 
total electric field can be shifted with respect to each other, 
$\vec{E}^+(x,t)=E_0e^{-i\nu t}[\vec{\epsilon}_+ \sin(kx+\theta)+\vec{\epsilon}_- 
\sin(kx-\theta)]$. We have denoted $k=\nu/c$ the laser wavevector, $E_0$ the 
amplitude and $\vec{\epsilon}_\pm$ unit right and left circular polarized 
vectors. The potentials ``seen'' by the atoms in the internal states $|0\r, 
|1\r$ are $V_0(x,\theta)=[V_{m_s=1/2}(x,\theta)+3V_{m_s=-1/2}(x,\theta)]/4$ and 
$V_1(x,\theta)=V_{m_s=1/2}(x,\theta)$, where $V_{m_s=\pm1/2}(x,\theta)=\alpha 
|E_0|^2 \sin^2(kx \pm \theta)$ \cite{Ja98}. 

This basic architecture can be used for quantum state control of neutral atoms 
in optical lattices \cite{Je01} and it constitutes the basis of the proposals 
for quantum computation in such systems \cite{Ja98,Br99}. As in the proposal of 
Ref. \cite{Ja98}, we make use of the fact that a relative movement of the two 
lattices, i.e. the trapping potentials $V_0,V_1$, can be achieved by varying the 
angle $\theta$.  In particular, starting with $\theta=0$, the respective minima 
of the potentials $V_0,V_1$ coincide and by changing $\theta$ from $0$ to 
$\pi/2$, the potentials $V_0,V_1$ move in opposite directions until their 
respective minima coincide again. Note that the shape of the potential $V_0$ 
changes as it moves.


\subsection{Implementation of the Hadamard walk}\label{implement}

We consider a single neutral atom at position $x_0=0$ and the case where lattice 
0 ---which traps the internal state $|0\r$ of the neutral atom--- moves with 
constant velocity to the left, $v_0=-v$, while lattice 1 ---which traps the 
internal state $|1\r$ of the atom--- moves with constant velocity $v_1=v$ to the 
right. The initial position of the lattices is such that the minimum of a potential well is 
located at position $x_0$ at $t_0=0$. The lattice movements are used to implement 
the controlled--shift operation (see Eq. (\ref{cshift})), while laser pulses 
allow one to manipulate the internal state of the atom and thus to select the the 
corresponding trapping potential (and therefore the direction of the movement).

Given that the atom is initially prepared in state $1/\sqrt{2}(|0\r+i|1\r)$ at position $x_0=0$, 
the application of the Hadamard operation (see Eq. (\ref{Had})) to the internal 
state of the atom at times $t_n=n d/v$ readily implements the quantum random 
walk on a line using this set--up. The spatial probability distribution of the atom at time 
$t_n$ ---i.e. the probability to observe an atom at position $kd, -n\leq k\leq 
n$ at time $t_n$--- corresponds exactly to $p(n,k)$ (Eq. (\ref{probHad})) of the 
one--dimensional Hadamard walk after $n$ steps. A simple fluorescence 
measurement ---together with several repetitions of the experiment--- allows one to 
measure this distribution.

To justify this statement, note that a single atom initially at position $x_0=0$ 
which is prepared in state $|0\r$ [$|1\r$] moves with constant velocity to the 
left, $x(t)=x_0-vt$ [right,  $x(t)=x_0+vt$] respectively. After a time $t_n=n 
d/v$, the position of an atom is shifted by exactly $n$ lattice periods $nd$ and 
the two lattices are again on top of each other. By changing the internal state 
of the atom e.g. at time $t_n$ from $|0\r$ to $|1\r$, one can switch between the 
corresponding trapping potential and thus change the direction of the movement. 
Note that it is important to make such changes of the internal state of the atom 
only when the two lattices are on top of each other, to ensure that the atom 
remains trapped in one of the potentials. Coherent superpositions of two 
internal states behave likewise. In the case of Rubidium with $|0\r = 
|F=1,m_f=1\r$, $|1\r = |F=2,m_f=2\r$, one can use standard Raman pulse or microwave 
techniques to realize the Hadamard rotation by using fast laser pulses. 

Note that on the Blochs sphere, the Hadamard operation corresponds to a rotation 
of an angle $\pi$ around the axis $\vec{u}=1\sqrt{2}(\vec{e_x}+\vec{e_z})$. This 
corresponds to a $\pi$--pulse rather than a $\pi/2$--pulse in the usual 
terminology of quantum optics. Up to an irrelevant global phase, one may also 
achieve the Hadamard operation by a sequence of three $\pi/2$--pulses, $H\propto 
e^{-i\pi/4\sigma_x}e^{-i\pi/4\sigma_y}e^{-i\pi/4\sigma_z}$. Experimentally it 
may be easier to use a $\pi/2$--pulse corresponding to the transformation 
$U_{\pi/2}=e^{-i\pi/4\sigma_x}$ instead of the Hadamard operation and prepare 
the atom initially in state $1/\sqrt{2}(|0\r+|1\r)$. This also leads to a 
symmetric probability distribution for all times $t$ equivalent to the one 
resulting from the standard Hadamard walk.

The Hadamard operation has to be applied at all lattice sites, which can be 
easily achieved by using a non--focused laser beam. In fact, such a homogeneous 
operation $H^{\otimes N}$ is much easier to implement than individual operations 
on specific lattice sites. This is due to the fact that in current experiments, 
the lattice period $d\approx 425$nm ---which is limited by the optical 
wavelength--- is smaller than the best achievable focusing width of the laser 
beams, $w\approx 1\mu$m \cite{Blnote}. In the fluorescence measurement, one can 
either detect unselectively both internal states $|0\r$ and $|1\r$ to reveal 
information about the position of the atom, or one may use selective resonance 
fluorescence methods. In the latter case, addition application of a random 
$\sigma_x$ operation ($\pi$--pulse) before the measurement is required to remove 
the dependence on the internal state of the spatial probability distribution. 
Provided the atom was initially prepared in state $1/\sqrt{2}(|0\r+i|1\r)$, the 
probability distribution is symmetric when tracing out the internal state of the 
atom, however the distributions conditioned on the internal state of the atom 
are asymmetric and have mirror symmetry, which explains the additional 
application of a random $\sigma_x$ operation. Notice the phase $i$ in the 
initial state, which is important to ensure symmetric behavior of the random 
walk.

We would like to emphasize that the procedure sketched above is readily 
implementable with existing technology. It does not require addressability of 
individual lattice sites. 

The essential requirements are that the internal states of the atom ---as well 
as their coherences--- are sufficiently stable and that the particle remains 
trapped in the potential throughout the procedure. This can be satisfied if the 
movement of the lattice is sufficiently slow [that is $v \ll \nu_{\rm osc}$, 
where $\nu_{\rm osc} \approx a_0\omega$ is the rms velocity of the atoms in the 
vibrational ground state, $\omega$ is the excitation frequency and $a_0$ is the 
size of the ground state of the trap potential \cite{Ja98}] such that the atom 
stays in the ground state of the potential during motion. This condition can be 
relaxed, as will be discussed in the subsequent section, and one can also allow 
non--adiabatic velocity profiles. The coherence of the internal state is mainly 
affected by fluctuations the intensity and phase of the trapping lasers 
\cite{Po99} as well as magnetic field fluctuations which may lead to 
uncontrollable energy splittings between the internal levels. We will address 
some of these issues in the next section. Given that these noise sources can be 
controlled sufficiently well, the number of steps of the random walk one can 
perform is only limited by the spontaneous emission lifetime of the atom in the 
lattice, which is at the order of several seconds. 
This corresponds to a maximum number of about $n=10^4$ time steps, assuming $t_1 
\approx 100\mu$s$-1$ms which respects the adiabaticity requirement for lattice 
shifts. 
Note that the implementation of several hundred time steps of the 
Hadamard walk corresponds to a spatial width of the quantum distribution at the 
order of millimeters.


\subsection{Improved implementation of the Hadamard walk}\label{improved}

From a practical point of view, there are a number of difficulties with the procedure 
proposed in Sec. \ref{implement}. For example, the laser pulses to 
implement the Hadamard rotation have to be fast compared to the timescale of the 
lattice movement. In addition, if the internal state of an atom is changed e.g. 
from $|0\r$ to $|1\r$, this implies a sudden momentum change of the atom, as it 
is no longer trapped in the left--moving lattice but in the right--moving one. 
This momentum change may lead to heating of the atom, and the atom may 
eventually even escape from the trap. 

Another practical difficulty one faces in current experiments is concerned with 
dephasing of the internal states of the atom. In particular, uncontrollable 
time and space dependent magnetic fields lead to energetic shifts of the 
internal levels, which result in relative phase shifts destroying the 
coherence of the system \cite{noteB}.

In this section, we propose a slight modification of the implementation 
suggested in Sec. \ref{implement}. This scheme is based on symmetrizing the 
procedure and avoids the problems mentioned above. Instead of moving the 
lattices with constant velocity, they oscillate around the central position $x_0=0$.
In the simplest case, the movement of the lattices is harmonic and may be described as follows
\bea
x'(t)&=&d/2(\cos{\omega_L t} -1), \nonumber\\
x''(t)&=&-d/2(\cos{\omega_L t} -1),
\eea
where $x'(t)$ [$x''(t)$] is the position of the central potential well of 
lattice 0 [1] respectively. The oscillation frequency $\omega_L$ is chosen in 
such a way that the adiabaticity requirement for the lattice movement ---i.e. 
that the atom remains in the motional ground state throughout the procedure--- 
are well fulfilled, which leads to oscillation times at the order of $100 \mu$s 
to ms. Note that one may replace the simple harmonic movement of the lattices 
by more complicated profiles, either specially suited to meet adiabaticity 
requirements such as the one proposed in Ref. \cite{Ja98}, or specially designed 
in such a way that adiabaticity requirements need not be matched, but are 
replaced by the weaker condition that the atoms ---after moving the lattice by one period--- are again in the motional ground state \cite{Po99}. These 
specially designed profiles may allow for movement times at the order of 
a few tens of microseconds for a shift of one lattice period. The profiles of Ref. 
\cite{Ja98,Po99} need to be adopted in such a way that the lattices oscillate 
around a central position, which may be accomplished by choosing the original 
profile until the lattice is displaced by one lattice period and the velocity is 
zero to good approximation, and then use the time--inverse profile. In this way, 
periodic lattice movements are readily achieved. In what follows, we will 
restrict our discussion to harmonic lattice movements, which is sufficient to 
illustrate the ideas of the improved implementation of the Hadamard walk. This is realized 
if, in addition, $\sigma_x$ operations ($\pi$--pulses) are applied at 
times $t_n\equiv n\pi/\omega_L, n=1,2,3,\ldots$ to all lattice sites, with the  
effect $|0\r \leftrightarrow |1\r$.

Under these conditions, an atom initially in state $|0\r$ is trapped in lattice 
0 and starts moving to the left together with lattice 0, where at time $t_1$ it 
is located at position $-d$ and the internal state changes to $|1\r$ due to the 
application of $\sigma_x$. The particle is therefore now trapped in lattice 1, 
where which moves to the left in the interval $(t_1,t_2)$, leaving the 
trapped atom at position $-2d$ at time $t_2$ etc.. On the one hand, we have that 
at times $t_n$ ---when manipulations of the internal states of the atoms are 
performed--- the two lattices are on top of each other and the velocity of the 
atoms is zero, which overcomes the first difficulty mentioned above. One could 
in principle also stop the lattice movement at these times until manipulation of 
the internal states is achieved, which allows to drop the requirement that 
manipulation of the internal states of the atom have to be fast compared to the 
timescale of the lattice movement. On the other hand, since within a time span 
of $2t_1$ the atom is both for time $t_1$ in state $|0\r$ and for $t_1$ in state 
$|1\r$, relative phase shifts between internal states $|0\r$ and $|1\r$ become 
an irrelevant global phase shift. Also fluctuation of magnetic fields become 
irrelevant, provided the timescale of the fluctuations is much larger than 
$2t_1$ and the spatial variation is negligible within $2d$ \cite{noteB2}. These 
requirements are well fulfilled e.g. for the typical 50Hz background noise and 
oscillating times $t_1\approx 100\mu$s$-1$ms when using harmonic lattice 
movements.

The implementation of the Hadamard walk using this set--up is straightforward. 
After the $\sigma_x$ operations at times $t_n$, the Hadamard operation $H$ (see 
Eq. (\ref{Had})) is applied at times $t_n$ if $n$ even, while $H'=\sigma_x H 
\sigma_x$ is applied at times $t_n$ if $n$ odd. That is, at $t_0$ $H$ is 
applied, while at $t_n$ $H'\sigma_x=\sigma_x H$ [$H\sigma_x$] is applied if $n$ 
is odd [even]. The use of the operation $H'$ instead of $H$ results from the 
interchanged role of the internal states $|0\r$ and $|1\r$ for even/odd $n$. One 
readily checks that in this way, after time $t_n$, $n$ steps of the quantum 
random Hadamard walk are implemented, provided the internal state of the atom is 
$1/\sqrt{2}(|0\r+i|1\r)$. The additional application of $H,H'$ does not 
influence the symmetrization discussed in the previous paragraph. Fluorescence 
measurements can be performed as described in Sec. \ref{implement}.


\subsection{Bounded random walks and random walk on a circle}

Using above set--up one can also implement an one- or two--side 
bounded random walk on a line \cite{Am01,Ba02}. Such a bounded random walk is similar to its 
unbounded counterpart, however at a certain locations $x_1,x_2$ barriers are 
introduced and the walks ends once a particle reaches one of these locations. A 
one side bounded random walk may e.g. contain a barrier at $x_1=100$, while $x_2 
\rightarrow -\infty$. Such a barrier can in the optical lattice set--up for instance 
be implemented by shining a laser at a certain location $x_1$, which couples 
both state $|0\r$ and $|1\r$ to a fast decaying auxiliary level. In such a way, a 
(position) measurement of the atom projecting onto $P_1=|x_1\r_X\l x_1|$, is performed. 

A modification of the trapping geometry may also allow for the implementation of 
a random walk on a circle, following the ideas of a recent proposal by Burke et 
al. \cite{Nist}, using an evanescent field of a linear waveguide and a ring 
resonator for trapping and guiding atoms (see also Ref. \cite{Nist1}). In this 
way, the periodic trapping potential can be modified such that lattices sites 
are located on a circle, forming a regularly spaced pattern. Movements of the 
trapping potentials result in this case in a circular movement of the lattice 
sites and thus of the trapped particle. By means of Hadamard rotations together 
with lattice movements, a random walk on a circle could be implemented. 
Measurements and manipulations of the trapped atom can be performed in the same 
way as discussed in Sec. \ref{implement} and Sec. \ref{improved}. Note that in 
contrast to the quantum random walk on the line, the random walk on the circle 
can not be implemented using existing experimental set--ups but rather relies on 
a proposed scheme.


\subsection{Using 2D set--up to measure probability distribution of a 1D random walk}

In current experiments, two or three dimensional lattice arrays are used rather 
than 1D--arrays. In this case, four [six] interfering laser beams constitute the 
two [three] dimensional trapping potential. One can make use of such a set--up 
to directly measure the probability distribution of the one--dimensional random 
walk. Consider a two dimensional lattice, which may be loaded  from a BEC 
\cite{Bl01}. We assume that the 2D lattice is loaded in such a way that in 
one dimension, say $x$, only the central lattice site $x_0$ is occupied, while 
in the other dimension, say $y$, all lattice sites are occupied in a regular way 
with one atom per site. This can be accomplished, for example, by first realizing a Mott 
transition in a 3D optical lattice from a BEC to a Mott insulator state as in recent experiments of the 
Munich group \cite{Bl01}, following a theoretical proposal of Ref. \cite{Ja98a}. In a 
second step, one can e.g. use of a gradient magnetic field to selectively address 
specific lattice layers and to deplete unneeded lattice sites \cite{Blnote}. One may also 
apply methods similar to the ones used in Ref. \cite{De99} to achieve uniform 
filling factors.

Such a configuration allows for a parallel sampling of one dimensional 
random walks (in the $x$ direction), by moving the lattices in the $x$ direction 
only and applying Hadamard rotations to all atoms as described in Sec. 
\ref{implement} and Sec. \ref{improved}. Each of the atoms at position $y_k=k d$ 
independently performs a random walk. A fluorescence measurement, e.g. a 
projective picture along the y--axis, allows to directly measure the 
corresponding probability distribution $p(n)$ provided the number of lattice 
sites in $y$ direction is big enough. Otherwise, the required number of 
repetitions of the experiment to determine the probability distribution is decreased.

\section{Experimental imperfections}\label{imperfect}

Although existing experiments using optical lattice systems offer high 
accuracy in both coherent storage and manipulation of the atoms, different 
kinds of errors may influence the ideal evolution. These errors may disturb or 
even destroy typical features of the quantum random walk, such as linear 
spreading time.  In this section, we concentrate mainly on errors in the 
coherence of the internal states of the atoms. We thereby observe a transition from 
quantum mechanical to classical behavior of the random walk for increasing 
errors. We expect these kind of errors to constitute the dominating part in the 
experimental imperfections. On the one hand, errors in lattice movements may 
lead to motional excitations of the atom. If sufficiently small these  
should however not effect the essential behavior of the system. On the other 
hand, the internal states of the atom are influenced by decoherence resulting from e.g. 
uncontrollable phase shifts, imperfections in the manipulation by means of laser 
pulses as well as fluctuations in the trapping potential during lattice 
shifts. One may distinguish between 
errors introduced by manipulations of internal state of the atom and errors in 
the position space of the atom, e.g. introduced by tunneling of atoms between 
neighboring lattice sites. While the former always keeps the structure of the 
ideal Hadamard walk that after $n$ time steps, only even [odd] lattice sites are 
occupied if $n$ is even [odd] respectively, errors in the latter lead to 
occupations of {\it all} lattice sites.

We use 
two simple models to investigate imperfections in the coherence of the internal 
state of the atom. We treat decoherence effects and errors in the manipulation 
of the atom (imperfect Hadamard operations), i.e. operations acting on the total 
Hilbert space ${\cal H}_I\otimes {\cal H}_X$ as $U_I \otimes \eins_X$, in a 
joint way. In the first model, we assume that the desired manipulation of the 
internal states of the atom, $U$, is performed with probability $p$ at each time step, while with 
probability $1-p$ a completely depolarized, random state is produced. The 
parameter $p$ serves not only as a measure of the accuracy of the operation 
---where $p=1$ describes perfect operations, while $p=0$ corresponds to a 
completely random operation--- but also includes other decoherence effects due 
to storage errors or phase fluctuations as well as lattice movements. Such a 
covariant error model reflects our limited knowledge about the specific type of 
error which occurred in the system. This error model has also been used in other 
contexts \cite{Br98} and is described by the following mapping 
\be
\label{mod1}
{\cal E}(\rho) = p U_I \rho U_I^\dagger + (1-p) 1/2 \eins_I\otimes \tr_I (\rho).
\ee
Note that this model is equivalent to a (partially) depolarizing channel, ${\cal 
E}(\rho)=p U_I\rho U_I^\dagger + (1-p) 1/4 \sum_{k=0}^3 \sigma_k^{(I)} \rho \sigma_k^{(I)}$, 
where $\sigma_k$ are Pauli matrices with $\sigma_0\equiv \eins$.

The second model only includes phase errors and is motivated by the expectation 
that phase fluctuation may be the dominating part of errors occurring in optical 
lattice systems. This error model is described by the following mapping
\be
\label{mod2}
{\cal E'}(\rho) = p' U_I \rho U_I^\dagger + (1-p') U_I \sigma_3^{(I)} \rho \sigma_3^{(I)} U_I^\dagger.
\ee

Note that if the optical potentials are not very deep, tunneling between 
neighboring sites may occur as well. We have used a simple model of incoherent 
tunneling --affecting only the position of the atom--- which is given by the 
following mapping \cite{noteTunnel}
\be
\label{mod3}
{\cal E''}(\rho) = q \rho + (1-q)/2 (U_+ \rho U_+^{\dagger} + U_- \rho U_-^{\dagger}),
\ee
where $U_{\pm}=\eins_I \otimes \sum_k |k\pm1\r_X\l k|$ is the unitary shift 
operator which moves the particle either one position to the left, $U_-$ or 
right, $U_+$. That is, with probability $q$ nothing happens ---and thus in total 
the desired evolution occurs---, while with probability $(1-q)$ a tunneling of 
the atom to one of the neighboring lattice sites occurs. 

\begin{figure}[ht]
\begin{picture}(230,200)
\put(-5,10){\epsfxsize=230pt\epsffile[64 210 551 590]{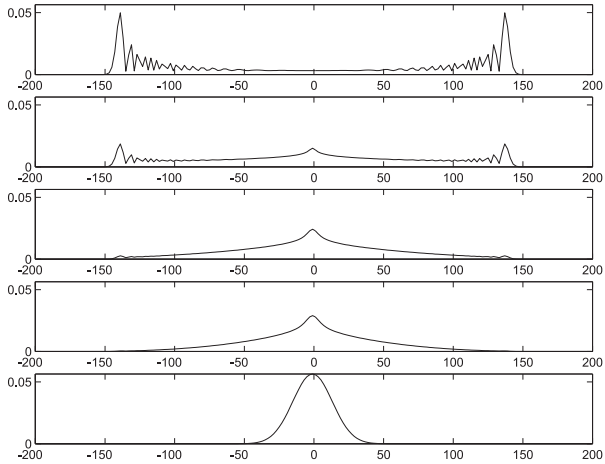}}
\end{picture}
\caption[]{Probability distribution after $n=200$ steps of quantum random walk on a line with imperfect operations using error model 1 (see Eq. \ref{mod1}). Error parameter $p=1, 0.99, 0.97, 0.95, 0$ from top to bottom. Only even positions are plotted, since odd positions are not occupied. The lowest curve corresponds to the probability distribution of the classical random walk on the line.}
\label{Fig1}
\end{figure}

\begin{figure}[ht]
\begin{picture}(230,200)
\put(-5,10){\epsfxsize=230pt\epsffile[64 210 551 590 ]{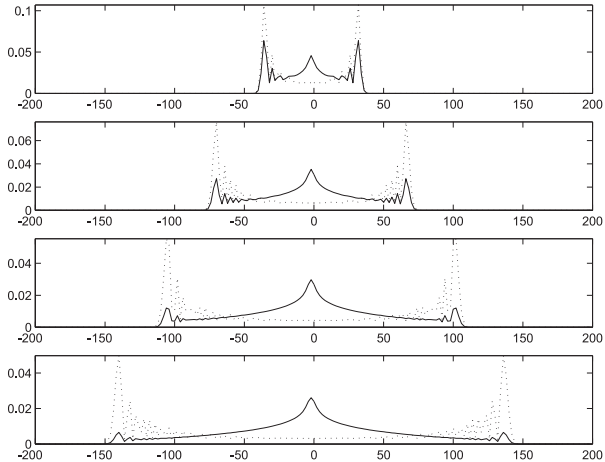}}
\end{picture}
\caption[]{Probability distribution for ideal quantum random walk on a line (dotted line) and random walk with imperfect operations using error model 2 (see Eq. \ref{mod2}) and $p'=0.98$ (solid line) after $n=50, 100, 150, 200$ steps (from top to bottom). Only even positions are plotted, since odd positions are not occupied. }
\label{Fig4}
\end{figure}

We have performed numerical simulations to investigate the influence of these 
kinds of errors on the quantum random walk on the line,  where we first assumed 
that errors affect only the internal state of the atom. Figure \ref{Fig1} is 
based on error model 1 and shows the probability distribution after $n=200$ 
steps of the random walk for different error parameters. Note that for 
completely random operations, i.e. parameter $p=0$, the particle performs 
exactly a classical random walk. This can easily be understood by observing that 
an internal state $1/2\eins=1/2(|0\r\l 0|+|1\r\l1|)$, when applying the 
controlled--shift operation (i.e. the lattice movement), has the effect that the 
particle moves with probability $1/2$ to the left, while with probability $1/2$ 
it moves to the right. In contrast to the quantum random walk, the resulting 
state is an {\it incoherent} superposition of the two possible states, which can 
be described classically and thus no interference effects (as in the quantum 
random walk) occur. The internal state plays the role of a classical coin. One 
observes from Fig. \ref{Fig1} that with increasing errors (decreasing parameter 
$p$), the probability distribution changes from the quantum mechanical one to 
the classical one. Even for errors at the order of several percent, typical 
quantum mechanical features of the probability distribution after a few hundred steps are clearly visible, 
in particular occupation in the interval $(\sqrt{n},n/\sqrt{2})$ can be 
observed. A similar simulation was performed using error model 2 (Eq. 
\ref{mod2}). The observed behavior of the system under this kind of error is 
very similar to the one shown in Fig. \ref{Fig1} using error model 1. Fig 
\ref{Fig4} shows the probability distribution after different number of steps of 
the ideal [imperfect] quantum random walk assuming only phase errors.

\begin{figure}[ht]
\begin{picture}(230,200)
\put(-5,10){\epsfxsize=230pt\epsffile[64 210 551 590 ]{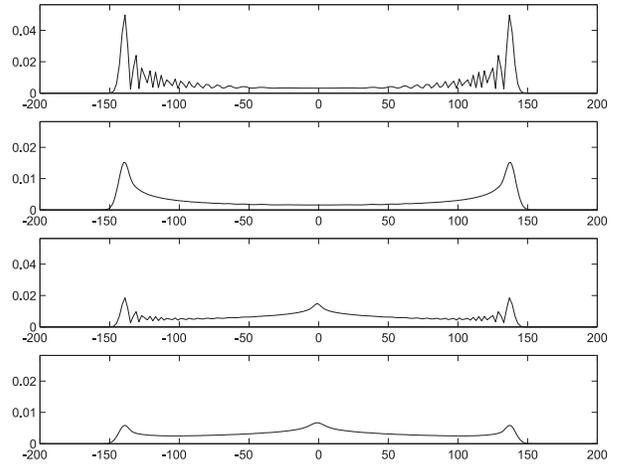}}
\end{picture}
\caption[]{Probability distribution after $n=200$ steps of quantum random walk on a line with imperfect operations using error model 1 (see Eq. \ref{mod1}) affecting internal state of the atom and errors due to tunneling described by Eq. (\ref{mod3}). Error parameters $(p;q)=(1;1),(1;0.95),(0.99;1),(0.99;0.95)$ from top to bottom. For $q=1$ only even positions are plotted, because odd positions are not occupied. For $q<1$, even and odd position are occupied and plotted, which explains the lower occupation of specific sites as compared to $q=1$.}
\label{Fig2}
\end{figure}

In the following we assume both internal and external errors (tunneling), described by 
Eq. (\ref{mod1}) and Eq. (\ref{mod3}) respectively. As shown in Fig. \ref{Fig2}, the 
essential effect of incoherent tunneling is that the probability distribution is smeared 
out. 

\begin{figure}[ht]
\begin{picture}(230,200)
\put(-5,10){\epsfxsize=230pt\epsffile[70 210 550 589]{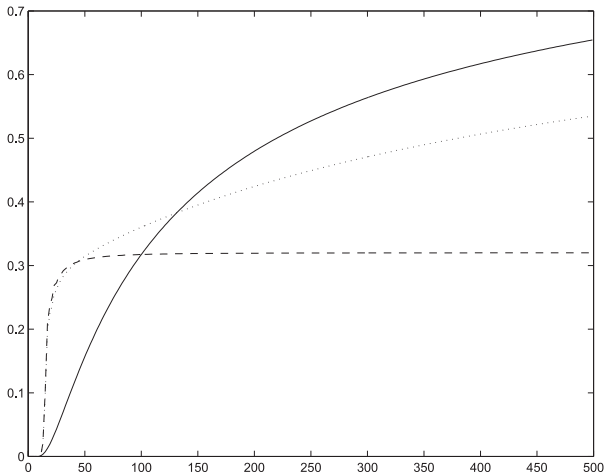}}
\end{picture}
\caption[]{Bounded random walk with barrier at position $x=-10$. Probability that atom was observed at the barrier plotted as a function of the number of steps for classical random walk (solid line), ideal quantum random walk (dashed line) and quantum random walk with imperfections in the manipulation of the internal state of the atom (dotted line), using error model 1 and $p=0.99$ (see Eq. \ref{mod1}).}
\label{Fig3}
\end{figure}

We have also considered a one--side bounded random walk. Fig. \ref{Fig3} shows 
the probability to observe the atom at the barrier at position $x=-10$ as a 
function of the number of steps. As expected, the exit probability for the 
quantum random walk is smaller than 1, while it approaches unity for the 
classical walk \cite{Ba02}.

Although no reliable estimates for the parameters $p$ including all possible 
imperfections and decoherence effects are available, errors at the order of 
several percent are still tolerable to observe a clear quantum behavior of the 
random walk even after a few hundred steps. This seems to be experimentally 
achievable. In turn, the distribution measured in the experiment can be used to 
determine the degree of coherence of the system, in particular the quality of 
the implemented operations. This may also serve as a test on the suitability of 
optical lattice systems to perform general purpose quantum computation, 
following the proposals of Ref. \cite{Ja98} and Ref. \cite{Ra01}. 


\section{Summary and conclusions}\label{summary}

We have proposed to use neutral atoms trapped in optical lattices to implement 
quantum random walks on the line and on the circle. The random walk is performed 
in position space by periodically shifting the lattices and manipulating the 
internal states of the atom(s) by homogeneous laser pulses. Read--out of the 
resulting probability distribution is performed via fluorescence measurements. 
Due to long life--times of the trapped atoms and efficient manipulation 
techniques, experimental realizability is expected with present--day technology. 
We have also investigated the influence of decoherence and imperfections in 
manipulation of internal state of the atoms and showed a transition taking place 
from the ideal quantum random walk to the classical random walk for increasing 
errors. Errors at the order of percent seem tolerable to still observe a clear 
quantum behavior of the walk after a few hundred steps.


\section*{Acknowledgements}
We would like to thank Immanuel Bloch and Markus Greiner for valuable 
discussions. H.-J.B. would like to thank Ignacio Cirac and Peter Zoller for 
useful comments. This work was supported by European Union through the Marie 
Curie fellowship HPMF-CT-2001-01209 (W.D.), IST-1999-13021 and the Deutsche 
Forschungsgemeinschaft. VK is funded by the UK Engineering and Physical Sciences 
Research Council grant number GR/N2507701.



\begin{thebibliography}{99}
\bibitem{noteB}
Depending on the energy $E_k \propto F |B(x,t)|$ of the hyperfine structure 
levels $F=1 (|0\r$) and $F=2 (|1\r$), we have that $|k\rangle \rightarrow e^{-i \phi_k}|k\rangle$, where $\phi_k \propto E_k t$.
\bibitem{noteB2}
Using that $B(x,t) \approx B_0$ within $2t_1$ and two lattice sites, we have 
that after a complete cycle (including also a final $\sigma_x$ operation), i.e. 
$2t_1$, $|k\r \rightarrow e^{-i(\phi_0+\phi_1)}|k\r$ which implies that relative 
phases became a global, irrelevant phase. For example, a particle in state 
$|0\r$ picks up a phase shift $\phi_0 t_1$ within the first time interval $(0,t_1)$, 
while a phase shift $\phi_1 t_1$ is picked up in the second time interval $(t_1,t_2)$.
\bibitem{noteTunnel}
We have restricted our investigation to incoherent tunneling effects. One may also investigate the influence of coherent tunneling on the quantum random walk. In this case, tunneling serves as additional source of interference which may affect the interference constituting the basis of the quantum random walk. A detailed analysis of such influence is however beyond the scope of this paper.
\end{thebibliography}
\end{document}